\documentclass[linenumbers]{aa}
\usepackage{graphicx} 
\usepackage{geometry}
\usepackage{ulem}
\usepackage{txfonts}
\usepackage{amsmath, amsfonts, amssymb}
\usepackage{xcolor}
\usepackage{caption, subcaption}
\usepackage{hyperref}
\usepackage{sansmath}

\usepackage{soul}
\DeclareRobustCommand{\VAN}[3]{#2}
\let\VANthebibliography\thebibliography
\def\thebibliography{\DeclareRobustCommand{\VAN}[3]{##3}\VANthebibliography}

\renewcommand{\emph}[1]{\textit{#1}}
\newcommand{\msun}{M$_\odot$}
\newcommand{\gcc}{g\,cm$^{-3}$}
\newcommand{\tbts}{$T_b$--$T_s$\xspace}

\begin{document}

   \title{Magnetar Outburst Models with Cooling Simulations}


   \author{D.\ De Grandis\thanks{\url{degrandis@ice.csic.es}}\inst{1,2}
          \and N.\ Rea\inst{1,2}
          \and K.\ Kovlakas\inst{1,2}
         \and F.\ Coti Zelati\inst{1,2}
         \and D.\ Vigan\`o\inst{1,2}
         \and S.\ Ascenzi\inst{3}
         \and J.\ A.\ Pons\inst{4}
         \and R.\ Turolla\inst{5,6}  
         \and S.\ Zane\inst{6}
          }

   \institute{Institute of Space Sciences (ICE-CSIC), Campus UAB, C/ de Can Magrans s/n, Cerdanyola del Vallès (Barcelona) 08193, Spain
     \and
             Institut d'Estudis Espacials de Catalunya (IEEC), 08034 Barcelona, Spain
    \and Gran Sasso Science Institute (GSSI), Viale F. Crispi 7, L’Aquila, 67100, Italy
    \and Departament de F\'isica Aplicada, Universitat d’Alacant, Ap. Correus 99, E-03080 Alacant, Spain  
    \and Department of Physics and Astronomy, University of Padova, via Marzolo 8, I-35131 Padova, Italy
    \and Mullard Space Science Laboratory, University College London, Holmbury St. Mary, Surrey, RH5 6NT, United Kingdom
}

   \date{Received XXX 00, 20XX; accepted YYY 00, 20YY}

 
  \abstract
   {Magnetar outbursts are one of the most noteworthy manifestations of magnetism in neutron stars. They are episodes in which the X-ray luminosity of a strongly magnetised neutron star swiftly rises by several orders of magnitude to then decay over the course of several months.
   In this work, we present simulations of outbursts as a consequence of localised heat deposition in a magnetised neutron star crust, and the subsequent surface cooling. In particular, we employ a magnetothermal evolution code adapted to the study of short-term phenomena, that is, one including in its integration domain the outer layers of the star, where heat diffusion is faster. This choice entailed the development and use of heat blanketing envelope models that are thinner than those found in the literature as the surface boundary condition. We find that such envelopes can support a higher surface temperature than the thicker ones (albeit for less time), which can account for the typical luminosities observed in outbursts even when coming from small hotspots (few km in radius).
   We study several parameters related to the energetics and geometry of the heating region, concluding that the cooling of a crustal hotspot found in the outer part of the crust can account for the luminosity evolution observed in outbursts both in terms of peak luminosity and timescales. Finally, we discuss the key observables that must be studied in future observations to better constrain the nature of the underlying mechanism.
   }


   \maketitle
%
\section{Introduction}
\label{sec:intro}

One of the most conspicuous traits of neutron stars (NSs) are their extremely strong magnetic fields, which are inferred to range from several millions up to $\approx10^{15}\,$G. At the higher end of this range are found the so-called magnetars, objects powered by their ultra-strong magnetic field; they manifest themselves as Soft Gamma Repeaters (SGRs) and anomalous X-ray pulsars (AXPs), though this distinction is by now largely obsolete \citep[see the reviews by \citealp{2015SSRv..191..315M, 2015RPPh...78k6901T, 2017ARA&A..55..261K, 2021ASSL..461...97E}]{1992ApJ...392L...9D}. Many magnetars have been observed to undergo outbursts, i.e.\ phases of enhanced X-ray luminosity that last from several months to years \citep{2011ASSP...21..247R, 2018MNRAS.474..961C}. Indeed, the discovery of new sources often occurs during an outburst. 
In some cases, the same source went through several outbursts separated by years of quiescence, e.g., most recently, {SGR\,1935$+$2154} \citep{2016MNRAS.457.3448I, 2024ApJ...965...87I}, XTE\,J1810$-$197 \citep{2021MNRAS.504.5244B}, {CXOU\,J164710.2$-$455216} \citep{2019MNRAS.484.2931B} and {1E\,1547$-$5408} \citep{2011A&A...529A..19B, 2020A&A...633A..31C}. Overall, $\sim30$ outbursts have been observed to date\footnote{\url{https://www.magnetars.ice.csic.es}}.

Typically, the enhancement of the luminosity is accompanied by an increased flaring activity, with the emission of several short bursts in the first few hours of the outburst as well as by timing anomalies such as glitches \citep[e.g.][]{2017ARA&A..55..261K}. Recently, the latest outburst of {SGR\,1935$+$2154} attracted renewed attention to these phenomena, as two short radio bursts were detected in temporal and spatial coincidence with it \citep{2020Natur.587...54C, STARE2, 2020ApJ...898L..29M}, reinforcing the idea of an association between magnetar outbursts and at least some fast radio bursts \citep[e.g.][but several alternative scenarios have been proposed, e.g. \citealp{2021ApJ...917...13S}]{2021Univ....7..453C, FRB_Review_2022}.
From a spectral standpoint, the X-ray emission of magnetars can generally be described as the combination of a thermal component (one or more blackbodies associated to the emission from the surface) plus a power law with photon index $\sim 2\text{--} 4$ (associated to processes in the magnetosphere). The flux enhancement during an outburst corresponds to a hardening of this spectrum that is typically well fit by a further blackbody \citep[e.g.][]{2011ASSP...21..247R}. The equivalent emission radius of this component is much smaller than that of the whole stellar surface (typically $\lesssim3\,$km). Concurrently, the pulsed fraction rises substantially. This provides evidence that outbursts are associated with the appearance of overheated regions in the NS crust. Moreover, the timescale for heat diffusion in the outer crust (where densities are lower than the neutron drip one $\rho_\text{ND}\approx4\times10^{11}\,$\gcc) matches that of the duration of outbursts.

These factors concur with the idea that outbursts are caused by some kind of abrupt event liberating a large amount of heat in or just above the NS crust, likely magnetic in origin. The exact nature of this trigger is still debated, in particular whether it originates from within the crust or it is the result of the activity in the magnetosphere \citep[e.g.][]{2009ApJ...703.1044B, 2019MNRAS.484L.124C}. In particular, in the former case the energy release is thought to follow a star-``quake'' caused by the accumulation of magnetic stress due to e.g.\ the Hall effect \citep[e.g.][and references therein]{2020ApJ...902L..32D} or a MHD instability related to it \citep{2019PhRvR...1c2049G}, possibly mediated by a thermoplastic wave \citep{2016ApJ...833..189L, 2023ApJ...947L..16L}. In this scenario, outbursts amount to a substantial crustal heating and the subsequent cooling, and they can hence be studied within the formalism of magneto-thermal evolution of NS crusts.
This approach has been first employed by \defcitealias{2012ApJ...750L...6P}{PR+12}\citet[in the following \citetalias{2012ApJ...750L...6P}]{2012ApJ...750L...6P}, using an early model in axial symmetry. Subsequent works have improved on this approach by enhancing the detail of the microphysics treatment \citep[which is based on a 1D MESA model]{2017ApJ...839...95D} or expanding the scheme to three dimensions, albeit in a setup with a simplified description of the microphysics \citep{2022ApJ...936...99D}. In this work, we revisit and expand the 2D approach with a more robust code, including a state-of-the-art microphysics input and extending the integration domain to low densities in order to consistently treat short-term events; moreover, we pave the way for detailed 3D modelling with recently-developed numerical tools \citep{2023MNRAS.518.1222D, 2024MNRAS.533..201A}. In particular, in Sec.\ \ref{sec:methods} we outline the adaptations we made to the cooling code for the study of short-term phenomena, with a particular focus on the choice of the envelope models (i.e.\ the boundary condition for the thermal evolution equation); we then proceed in Sec.\ \ref{sec:results} to relate different observables to the parameters of our model, in particular those regarding the energetics (Sec.\ \ref{sec:einj}, \ref{sec:tinj}) and the geometry (Sec.\ \ref{sec:geometry}) of the heat source; we then discuss our findings and draw our conclusions in Sec.\ \ref{sec:conclusions}. 
\begin{figure}
\centering
\begin{subfigure}{.48\textwidth}
\includegraphics[width=\textwidth]{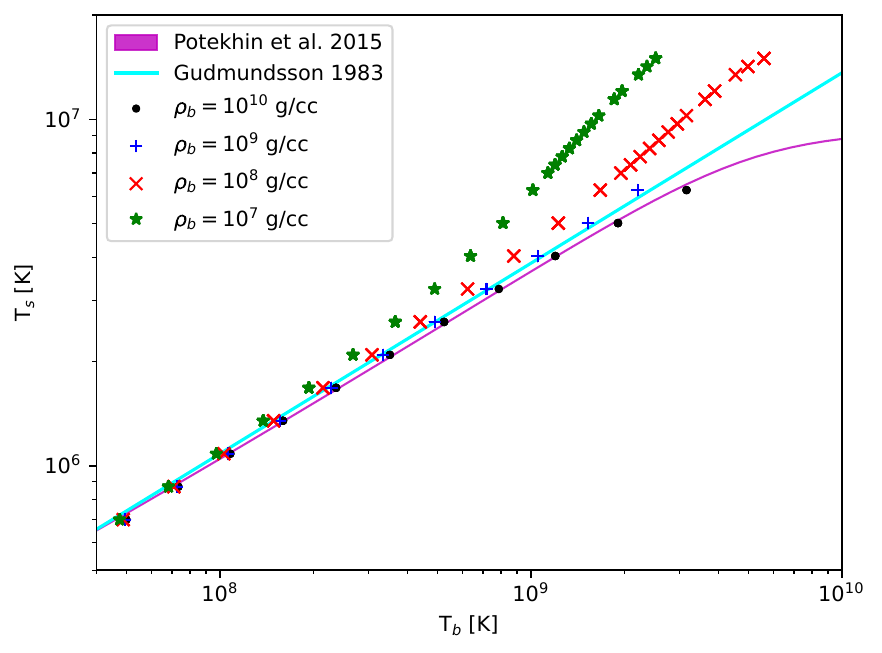}
\caption{$B=10^8\,$G}
\end{subfigure}
\begin{subfigure}{.47\textwidth}
\includegraphics[width=\textwidth]{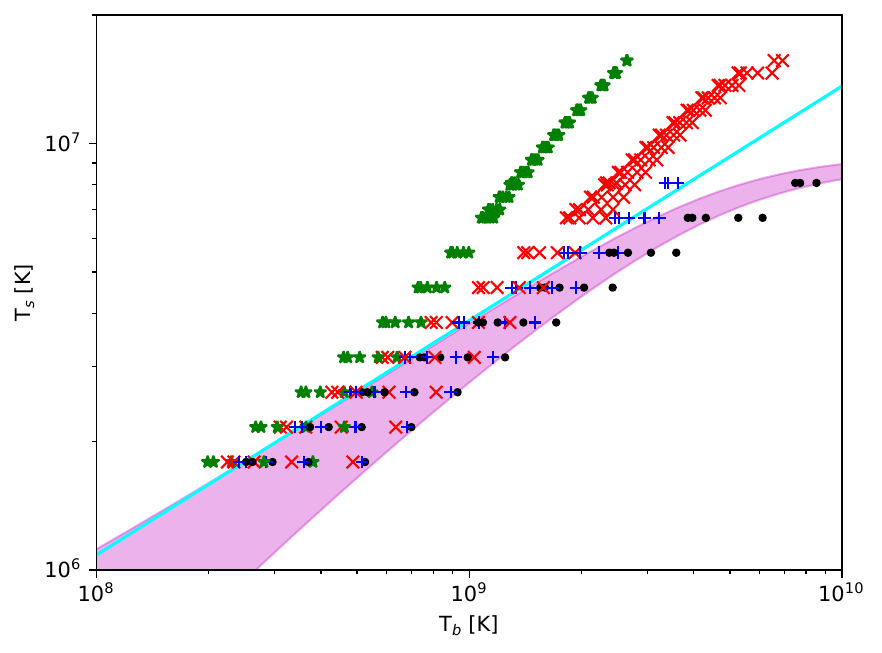}
\caption{$B=10^{12}\,$G}
\end{subfigure}
\caption{\tbts relations calculated for a sample of representative values of $T_s$ in the case of Fe envelopes. Different colours and markers denote different bottom-of-the-envelope densities $\rho_b$. The models from \citet{1983ApJ...272..286G} and \citet{transport}, calculated for the standard value $\rho_b=10^{10}\,$\gcc, are plotted for reference. The spread of the points at the same $T_s$ (which is noticeable only in the bottom panel) corresponds to different $\Theta_B$, with higher $T_b$ giving a lower $T_s$ value for the same $T_b$. Two values for the magnetic field, $10^8\,$G (top panel) and $10^{12}\,$G (bottom panel) are shown; these values have been chosen, although comparatively low, for clarity of visualisation, as stronger fields yield very sparse plots. }
\label{fig:tbts}
\end{figure}

\section{Methods}
\label{sec:methods}

\subsection{Physics input and numerical setup}
We employ a suitably adapted version of the 2D magnetothermal evolution code described in \citet{2021CoPhC.26508001V}, that solves the coupled thermal and magnetic evolution of a NS in an axisymmetric setup.
The hydrostatic structure of the star is calculated self-consistently solving the TOV equation \citep[e.g.][]{1977ApJ...212..825T} supplemented by the BSk24 equation of state \citep{10.1093/mnras/stz800} as found in the {CompOSE} catalogue \citep{Typel2015}\footnote{\url{https://compose.obspm.fr/eos/253}}.

In particular, the evolution of the temperature is found by solving the thermal evolution equation in the form
\begin{equation}\label{eq:heat}
    c_\text{v}\frac{\partial (\text{e}^\nu T)}{\partial t}=\nabla\cdot\left[\text{e}^{\nu}\,\hat\kappa\cdot\nabla(\text{e}^\nu T)\right] + \text{e}^{2\nu}(H-Q_\nu)
\end{equation}
where $c_\text{v}$ is the specific heat, e$^\nu$ is the general-relativistic metric factor calculated assuming a Schwarzschild metric (the operator $\nabla$ also contains a relativistic correction), $\hat\kappa$ the thermal conductivity tensor, $H$ is a heating term and $Q_\nu$ the neutrino emissivity. The term $H$, is akin to the one included in long-term cooling simulations to describe e.g.\ the heating coming from ohmic dissipation. In this instance, we activate it in a localised region of the crust (see Sec.\ \ref{sec:results}) and a short time period as a means to describe the heating associated to the activation of an outburst, without exactly specifying the nature of the trigger.

We adopted calculations of the transport coefficients described in \citet{transport} in their latest implementation available online\footnote{\url{https://www.ioffe.ru/astro/conduct/conduct.html}}. The neutrino emissivities are described according to \citet{2001PhR...354....1Y}; in particular, whereas the long-term cooling of a NS is controlled by the neutrino reactions happening in its core \citep{2004ApJS..155..623P}, when studying the thermal evolution of hotspots located in the crust it becomes fundamental to consider those happening in the crust itself. These are non-negligible only if the temperature is above $\gtrsim10^9\,$K, and hence are only important in the very early phases of ordinary cooling. However, their dependence on temperature is very steep, so that they are able to limit the maximum possible temperature that in a NS crust close to their activation threshold.

Within the present axially symmetric setup, we are only able to consider outbursts originating at one of the magnetic poles so that the heated region has a spot-like rather than a ring-like structure (which would alter the geometric relation between heated volume and emitting surface). The heating term $H$ in Eq.\ \ref{eq:heat} is therefore defined in terms of five parameters: the total energy input, $E_{\rm{inj}}$ the injection time $t_{{inj}}$, and three values defining the geometry of the heated region. Namely, we consider a truncated cone centred around the pole, defined by its radius $R$ (referred to the $z$ axis) and the inner and outer depths, counted from the crust-envelope interface (i.e.\ the external boundary of the grid), $z_{\rm{in}}$ (towards the centre) and $z_{\rm{out}}$ (towards the surface). 
In order to avoid numerically problematic sharp time gradients, the energy injection was modulated by a sinusoidal function in time; we tested other modulation profiles (e.g.\ a Gaussian), finding only very marginal differences in the shape of the luminosity curves near their peaks. 

In the present study, we do not consider the evolution of the magnetic field, but rather fix it in a configuration that for simplicity we chose to be dipolar, akin to the one described in \citet{2008A&A...486..255A}. 
Assuming no magnetic evolution over our simulations is reasonable since the magnetic field evolves over a timescales that are much longer that the months/years studied here. This assumption also amounts to disregarding any change in the field that is associated to the onset of the outburst itself, which reflects the choice of not considering a concrete trigger model (itself due to the lack of a viable picture in the literature). In the present setup, hence, the magnetic field acts as a bystander to the cooling of the hotspot. A more complete assessment of its role is beyond our present scope, as it would require both a sound microphysical treatment of the heating phase and a 3D framework (see Sec.\ \ref{sec:conclusions} for a more detailed assessment of the impact of the magnetic field on our results). Note that we keep the full dependencies of the microphysical quantities on the (fixed) magnetic field, most notably through the thermal conductivity and weak-synchrotron neutrino emission, even though it does not get evolved itself.
 
\begin{figure}
 \includegraphics[width=.47\textwidth]{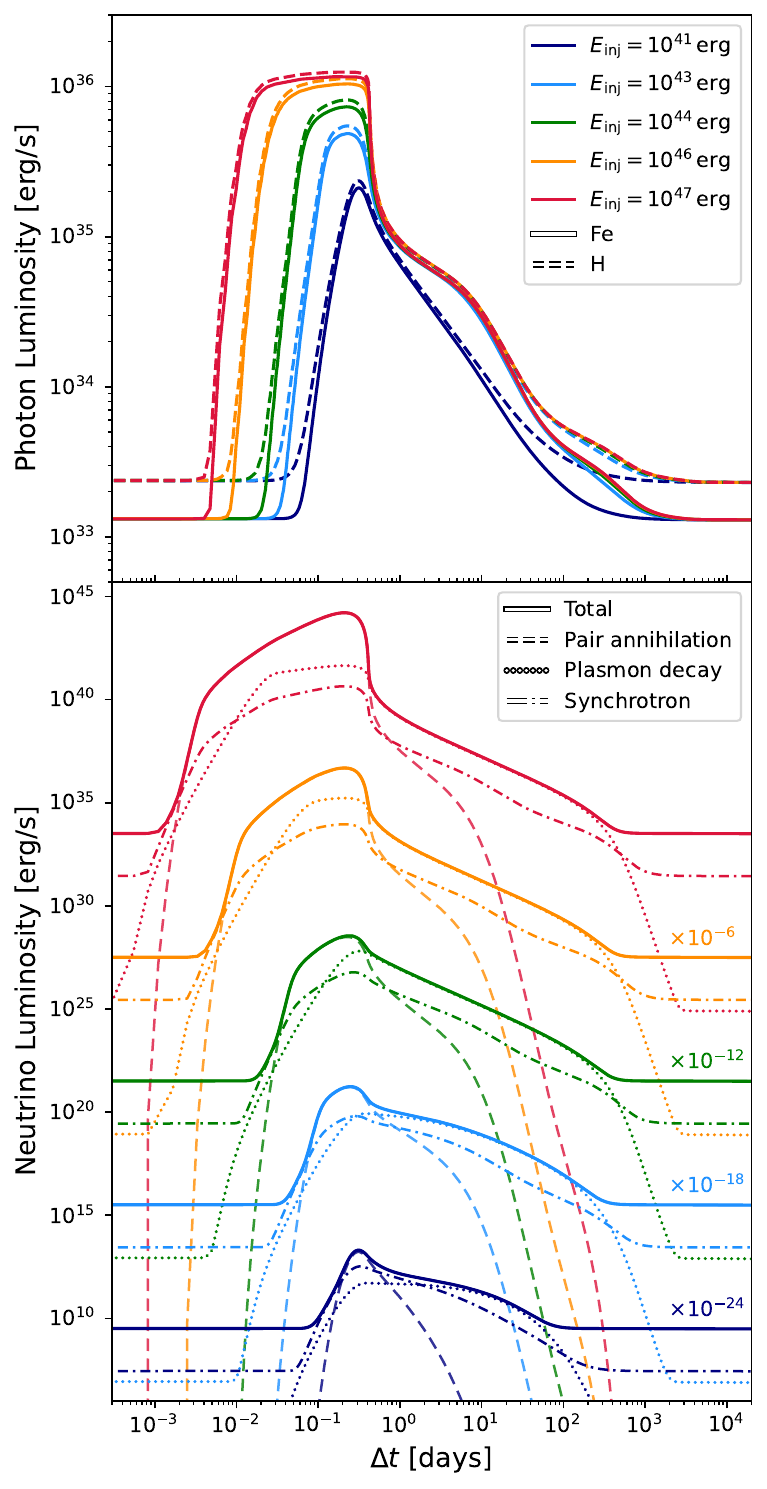}
    \caption{(\emph{Top panel})
    Photon luminosity with growing heat injection, in the case of Fe (solid lines) or H (dashed lines) envelopes. (\emph{Bottom panel}) The corresponding neutrino luminosity, with its dominant channels: a continuous line indicates the total $L_\nu$, dashed lines the contribution from e$^+$e$^-$ pair annihilation, dotted lines the one from plasmon decay and the dot-dashed the weak-synchrotron emission. The baseline of neutrino luminosity, not plotted separately, is in this case due to the breaking of $^2P_2$ Cooper pairs in the core. For clarity of presentation, we rescaled each family of curves by a factor indicated in the figure, and only the curves calculated with the H envelope are displayed (the ones corresponding to the Fe envelope are almost indistinguishable from them).
    }
    \label{fig:H}
\end{figure}

\subsection{Blanketing envelope models}

In the outermost stellar layers, density drops by several orders of magnitude within several tens of meters. The temperature also shows a steep gradient, dropping by about two orders of magnitude across this optically thick but geometrically thin layer. Hence, it is problematic to include this region in a computational domain together with the rest of the NS. Rather, it is standard practice to solve its thermal structure separately, to then use the result of this operation to set a Neumann-type outer boundary condition for the heat equation. Within this approach, the outermost layers are referred to as a heat blanketing envelope. In particular, an envelope model gets implemented as a relation between the temperature $T_b$ at the bottom of the envelope (i.e.\ at the top of the crust) and $T_s$ at the surface.

Several relations of this kind have been 
computed in the literature, the most well known being the one obtained by \citet{1983ApJ...272..286G} for an unmagnetised heavy-element envelope in plane-parallel approximation. Several alternatives, including ones accounting for e.g.\ different chemical composition or magnetic fields, are found in the literature \citep[see the review by][]{2021PhR...919....1B}. However, the vast majority of the \tbts relations available for application in cooling codes are adapted to the long-term evolution of NSs only. In fact, the assumption underlying the separation of the envelope from the rest of the integration domain is that the heat diffusion across the envelope itself should happen faster than the individual timestep of the simulation. Therefore, the thickness of the envelope sets the time resolution of the cooling simulation, and in order to study short-term phenomena it becomes necessary to consider a thinner envelope. At the same time, this operation moves a larger portion of the NS crust within the integration domain of the (in this case, 2D) magnetothermal code, so that the evolution of lower-density layers, where heat diffusion is faster, can be described with the full Eq.\ \ref{eq:heat} rather than in the plane-parallel, stationary approximation. We then proceed to build our own \tbts relations for envelopes that are thinner than those available in the literature taking into account the effects of a strong magnetic field.

We follow closely the formalism in \citet[see also \citealp{1977ApJ...212..825T}]{2007Ap&SS.308..353P}, producing a set of models of the outer stellar layers for different values of surface temperatures $T_s$, magnetic field strength and orientation of the field itself with respect to the radial direction. 
The integration of the corresponding system of differential equations proceeds up to a density $\rho_b$ that sets the thickness of the envelope. Whereas most models in the literature set $\rho_b=10^{10}\,$\gcc, we consider densities reaching down to $\rho_b\approx10^{7}\,$\gcc, so that the time resolution of the simulation goes down from the timescale of the year to that of the hour. These thinner envelopes are able to support larger values of the surface temperature for the same $T_b$, which cannot be sustained for a longer time in a stationary fashion. Some examples of how different choices of $\rho_b$ reflect on the \tbts relation are shown in Fig.\ \ref{fig:tbts}. Moreover, the maximum surface temperature is also controlled by neutrino emission in the envelope itself. This effect was described by \citet{transport} as an asymptotic limit to the \tbts relation (magenta shaded area in Fig.\ \ref{fig:tbts}). However, neutrino emissivity is significant only at densities above $\approx10^8\,$\gcc, so that for thinner envelopes this effect is not as important.

We built models of a thin ($\rho_b\approx10^{7}\,$\gcc), magnetised blanketing envelopes for two extreme cases of their chemical composition (Fe or H) and implemented them within our cooling code with a formalism that allows flexibility in $\rho_b$, as well as in $T_b$ and $\mathbf{B}$. Namely, we encapsulate our \tbts relation in a single-layer neural network, trained on a set of envelope models that replaces an analytical fit. We refer to a companion paper \citep{Kovlakas25} for details about this implementation, and about the construction of the envelope models themselves.

\section{Results}\label{sec:results}

In this section, we present a range of models of heat injection and subsequent cooling, varying the parameters associated to the heating itself, $E_\text{inj}$ and $t_\text{inj}$, and the geometry of the interested region $z_\text{in}$, $z_\text{out}$ and $R$. We will also address the role of the background state, in particular considering different NS masses.

\subsection{Increasing the energy input: luminosity saturation}
\label{sec:einj}

\citetalias{2012ApJ...750L...6P} observed that the peak luminosity of an outburst is capped by the steeply increasing efficiency of neutrino emissivity from the crust when the temperature gets above $\approx3\times10^9\,$K. This is in line with the observational fact that the luminosity of outbursts does not exceed a value of about $10^{36}\,$erg/s. However, the exact value crucially depends on the size of the heated region, as well as on the employed envelope model.

To reassess this result in our more consistent setup, we considered models with increasing $E_{\rm{inj}}$, for which we show in Fig.\ \ref{fig:H} the evolution  of photon and neutrino luminosity. Here, we kept a fixed injection time $t_\text{inj}=10\,$h and geometry, $z_{\rm{in}}=400\,$m, $z_{\rm{out}}=5\,$m, $R=3\,$km
; an analysis of the role of these other parameters follows in the next subsections. Indeed, we can observe that as $E_\text{inj}$ increases the photon luminosity curves get piled up at a limit value of about $10^{36}\,$erg/s. We tested two different envelope compositions, one of pure Fe and the other of pure H \citep[representing the two extreme scenarios, see][]{Kovlakas25}, finding just a small difference in the two cases. This is due to the fact that at very high temperature the thermal structure of the envelope itself is regulated by neutrino emission, so that the composition becomes less important.

On the other hand, neutrino emission keeps growing without reaching a limit value, even though the contributions from different weak reactions vary throughout the event. Namely, in the cases at hand three channels of neutrino production in the crust are relevant: in the early phases, when the temperature of the outer layers is at its highest, electron-positron pair annihilation dominates and is thus the main limiting factor for the rise of photon luminosity\footnote{The expression for this emissivity contribution used here is the one proposed by \citet{1996ApJS..102..411I}, which does not take into account the effect of the magnetic field, for simplicity of implementation. We tested as an alternative the one proposed in \citet{2001PhR...354....1Y}, finding an analogous behaviour; still, the literature lacks more modern, magnetised calculations.}. Later on, the cooling of the outburst is mainly regulated by weak plasmon decay, and for very strong fields (such as here, $B_d=10^{14}\,$G) synchrotron radiation can also get comparable to the other terms.

It must be kept in mind that the values of $E_\text{inj}$ studied here not directly comparable to those inferred in observational studies. This is due to the fact that observations are only able to access the energy fraction that is radiated as photons, whereas it cannot take into account that radiated as neutrinos. As the energy input increases, the latter contribution becomes more and more dominant, so that for total injections $\gtrsim10^{44}\,$erg the bulk of the emitted energy might go undetected. Still, the highest values considered in Fig.\ \ref{fig:H} are actually comparable to the total energy budget of a magnetar, and must therefore be taken as extremal cases. In the following, we adopt a more realistic value $E_\text{inj}=10^{43}\,$erg as a demonstrative value.

\subsection{Dependence on the injection time}\label{sec:tinj}
\begin{figure*}
    \centering
    \includegraphics[width=.9\textwidth]{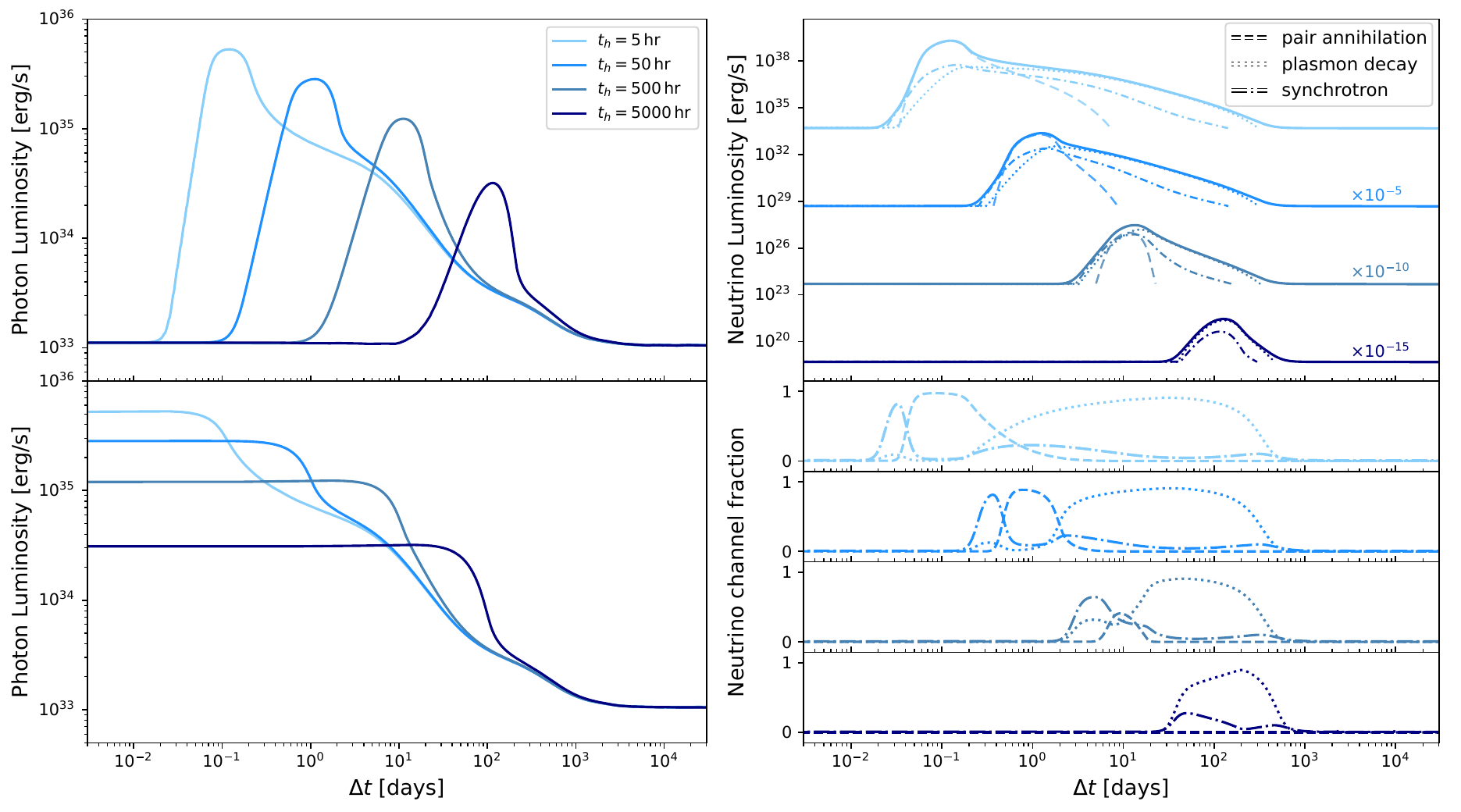}
    \caption{(\emph{Left}) Evolution of the photon luminosity varying the time $t_\text{inj}$ for which the injection term is active while keeping the other parameters fixed (though not visually obvious due to the log scale, the total area under each set of curves is the same). The lower panel shows the same curves as the upper one, but with $\Delta t=0$ corresponding to the time of the peak rather than the time of heating onset to more directly compare to the curves displayed in observational studies (see text). (\emph{Upper right}) Corresponding neutrino luminosity, rescaled for visualisation as in Fig.\ \ref{fig:H}. For clarity, the contribution of the neutrino processes are shown separately only when they are above the quiescence level. (\emph{Lower right}) Fraction of the total neutrino luminosity coming from the different crustal neutrino reactions that get activated during the outbursts.}
    \label{fig:th_LL}
\end{figure*}
Given the overall duration of an outburst, it is generally accepted that the heat dissipation mechanism triggering it should be quite fast, although there is no consensus on the exact timescale. An observational constraint can be put in the rare cases in which a magnetar had been serendipitously observed shortly before it went into an outburst, which in the most fortunate cases happened few days before the event \citep{2006ATel..902....1M, 2008MNRAS.390L..34E, 2013ApJ...770L..24K, 2017ApJ...847...85Y}.

The detection of several short X-ray bursts associated to the first hours of some of the events \citep[e.g.][]{2004ApJ...607..959G, 10.1111/j.1745-3933.2007.00317.x, 
2009ApJ...696L..74M,2010ApJ...711L...1V,2020ApJ...904L..21Y} may suggest a mechanism operating on such a timescale. On the other hand, the dissipation of thermoplastic waves in NS crusts is predicted to happen over a span of days to weeks, whereas the healing of crustal lattices following a magnetically-induced failure is estimated to take $\lesssim1\,$yr \citep{2016ApJ...833..189L}. 

Given this large uncertainty, and considering that there may be different mechanisms, and hence timescales, at play, we tested a range of heat injection times $t_\text{inj}$ spanning from a few hours to a few years, as shown in Fig.\ \ref{fig:th_LL}. The peak follows the length of the injection, whereas the maximum luminosity gets somewhat reduced for longer $t_\text{inj}$, as more energy is radiated via neutrinos when the heating is more spread in time; note also that in the latter case the values of temperature reached inside the crust are lower, so that the cooling is mainly controlled by weak plasmon decay rather than by the electron-positron weak annihilation that caps heating at higher temperatures.

In order to display more clearly the cooling curves, we use as the independent variable the time difference with respect to the beginning of the injection $\Delta t$, which is however an information that is not available for observational data. Conversely, to present a cooling curve that is more akin to what can be observed, in the second panel of Fig.\ \ref{fig:th_LL} we plot the curves computing $\Delta t$ from the time of the peak (which does not amount to a visual translation when using a log-scale). This operation removes a lot of information and blurs, to an extent, the differences between the lines, demonstrating how new data about the rise phase would be fundamental in constraining $t_\text{inj}$.

\subsection{Dependence on the injection region geometry}
\label{sec:geometry}
\begin{figure}
\includegraphics[width=.47\textwidth]{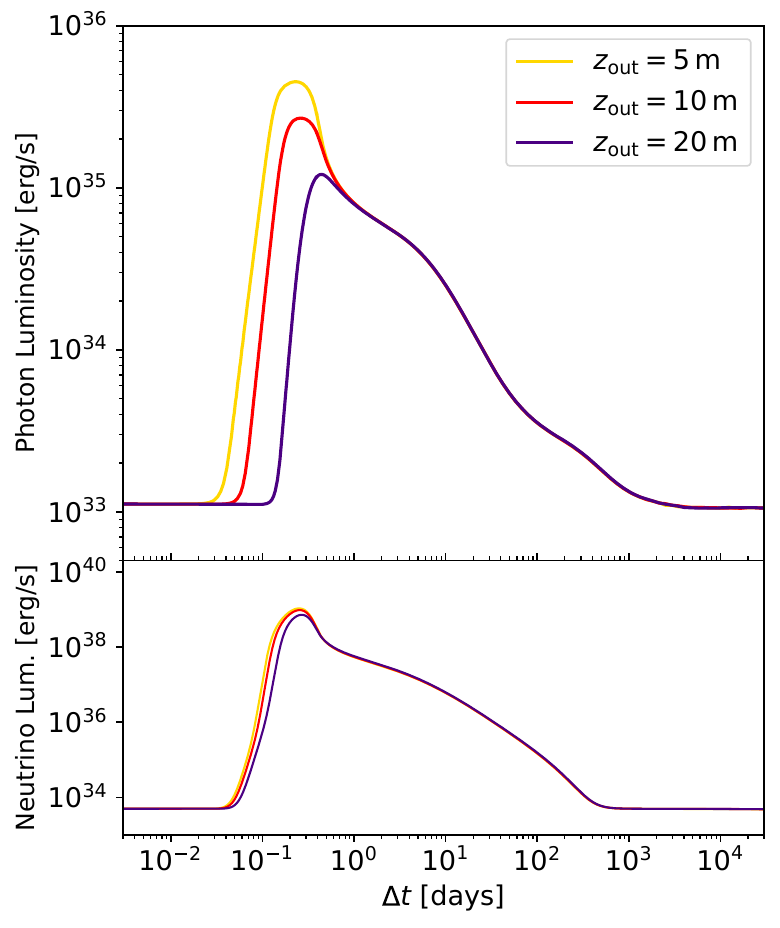}
    \caption{Same of Fig.\ \ref{fig:th_LL}, varying the outermost limit of the heated region $z_{\text{out}}$ as reported in the caption, keeping all other parameters fixed, for two envelope models. In this case, the separate contributions are not shown, as they only show marginal differences.}
    \label{fig:rout}
\end{figure}

\begin{figure}
    \centering
    \includegraphics[width=.47\textwidth]{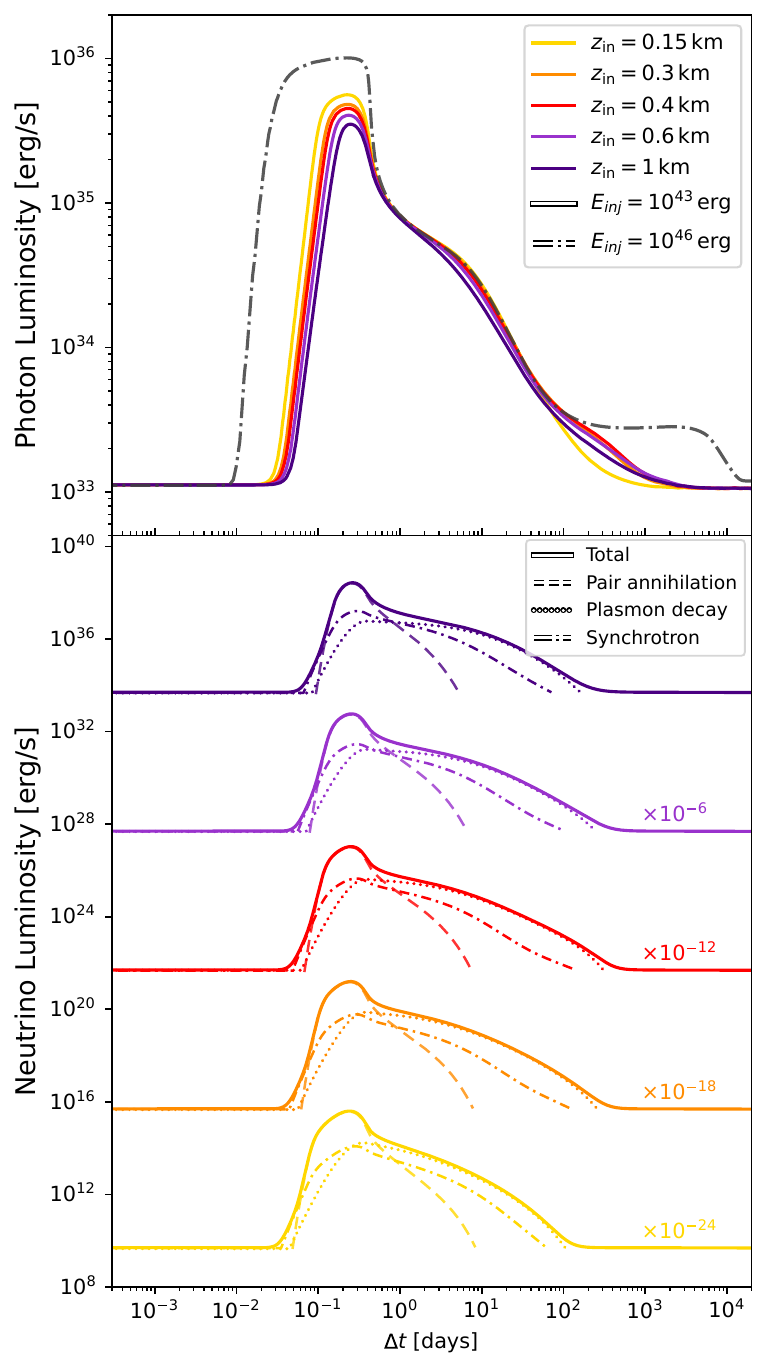}
    \caption{Same as Fig.\ \ref{fig:th_LL} varying the inner limit of the injection $z_{\text{in}}$ as reported in the caption. In the cases with the largest $z_\text{in}$, i.e.\ with the heating reaching higher density layers, the contribution of two further neutrino emission channels becomes important, namely the emissivity from 
    the formation/breaking of Cooper pairs of superfluid neutrons \citep[e.g.][]{2009ApJ...707.1131P} (empty-dotted lines) and the one from n-n and e-ion bremsstrahlung (double dashed lines).}
    \label{fig:rin}
\end{figure}

We now consider the effects of the geometry of the heated region on the cooling of the surface hotspot. First, we ran several cases varying the upper injection depth $z_{\text{out}}$, shown in Fig.\ \ref{fig:rout}, and the lower one $z_{\text{in}}$, shown in Fig.\ \ref{fig:rin}. Overall, these two parameters control the early and late phase of the event, respectively, as the heat diffusion time from to the surface from hot layers increases with their depth.

The outer limit $z_{\text{out}}$ also strongly affects the peak luminosity: in fact, at lower densities the heat capacity is lower, and a quicker diffusion the radiating surface reduces the amount of energy that is radiated as neutrinos before it can contribute to the luminosity at the surface. In the case of a very shallow injection, $z_{\text{out}}=5\,$m with a realistic hotspot size $R=3\,$km (see Sec.\ \ref{sec:geometry}), we obtain a peak value $L_{\text{max}}\sim10^{36}\,$erg/s. This is consistent with outburst observations, which show peak luminosities in the $10^{34}-10^{36}\,$erg/s range \citep{2018MNRAS.474..961C}.

Conversely, the inner depth $z_\text{in}$ affects the profile of the outburst at later phases. In particular, as the heat is deposited in denser and denser layers a shoulder appears in the late-time light curve, reflecting an increasing time for heat to diffusively reach the surface. 
However, this trend seems to be reversed in those  cases in which $z_{\text{in}}$ is large enough for heat to be deposited underneath the neutron drip point (in our case, located at depth $z_\text{drip}\simeq0.8\,$km), as the shoulder recedes. This is due to the discontinuity of the microphysical properties of the crust at that point, and in particular to the increased specific heat that renders the heat addition in the lower layers less effective. To further assess the effects of this discontinuity, we repeated the simulation with $z_\text{in}=600\,$m with a much larger heat input, $E_\text{inj}=10^{46}\,$erg, shown as the gray dot-dashed line in Fig.\ \ref{fig:rin}. In this case the deeper layers can experience a substantial temperature variation, and the heat emerging from these regions is delayed by the discontinuity in conductivity, forming a plateau-like structure at late times (which also enhances the time before returning to the quiescence luminosity by a substantial factor).

\begin{figure}
    \includegraphics[width=.47\textwidth]{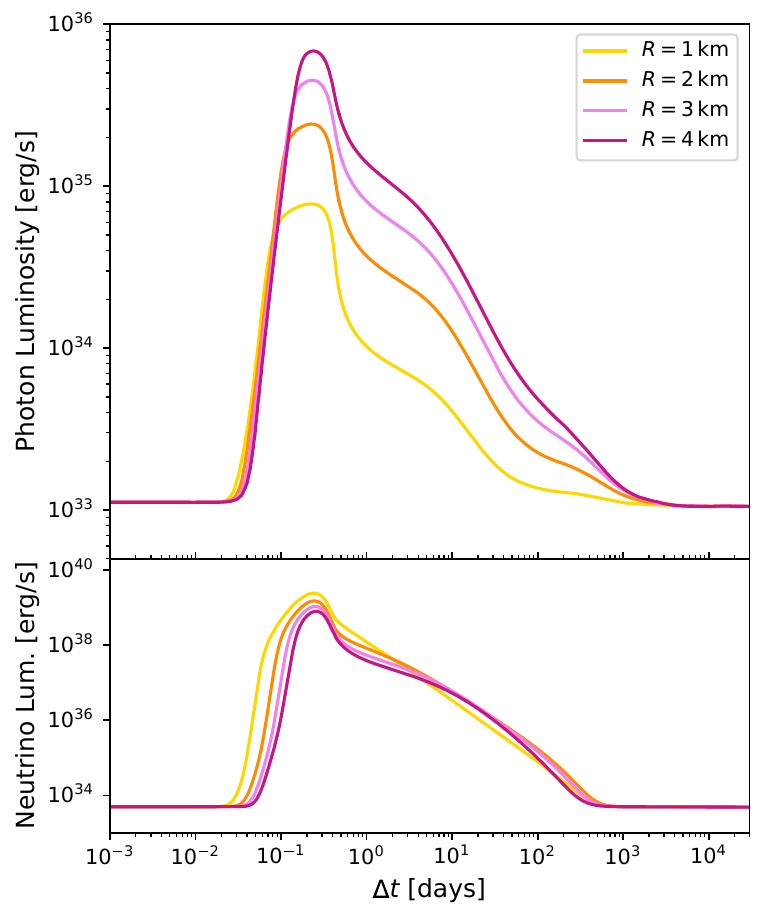}
    \caption{Same as Fig.\ \ref{fig:th_LL} varying the radius of the heated region $R$.}
    \label{fig:R_LL}
\end{figure}

Finally, in Fig.\ \ref{fig:R_LL} we show the effect of taking a different radius for the injection region. We chose four values for $R$, namely $1$, $2$, $3$ and $4\,$km, as the transient blackbody components characterising the spectrum of outbursts have equivalent emission radii of typical length $\lesssim3\,$km in the early phases, before shrinking during cooling. In the current setup, the different choices for the initial $R$ yield curves with the same traits, and almost amount to a rescaling of the outburst luminosity, i.e.\ to geometric effect due to a larger emission surface. It should be noted, however, that this length is related but not equal to the emission radius as obtained from observations, which is affected by general-relativistic ray-bending and the systematics associated to spectral fits. A more detailed study of these effects is deferred to future work. In our runs, we did not observe a marked evolution of the hot region, mainly due to the choice of a polar patch, in which the field lines are almost perpendicular to the surface and thus hinder heat diffusion. However, simulations studied in a 3D setup \citep{2022ApJ...936...99D} showed that the magnetic field can significantly alter the shape of the hotspot (see Sec.\ \ref{sec:conclusions}).

\subsection{The role of the background state: different stellar masses}

\begin{figure}
    \centering
    \includegraphics[width=.49\textwidth]{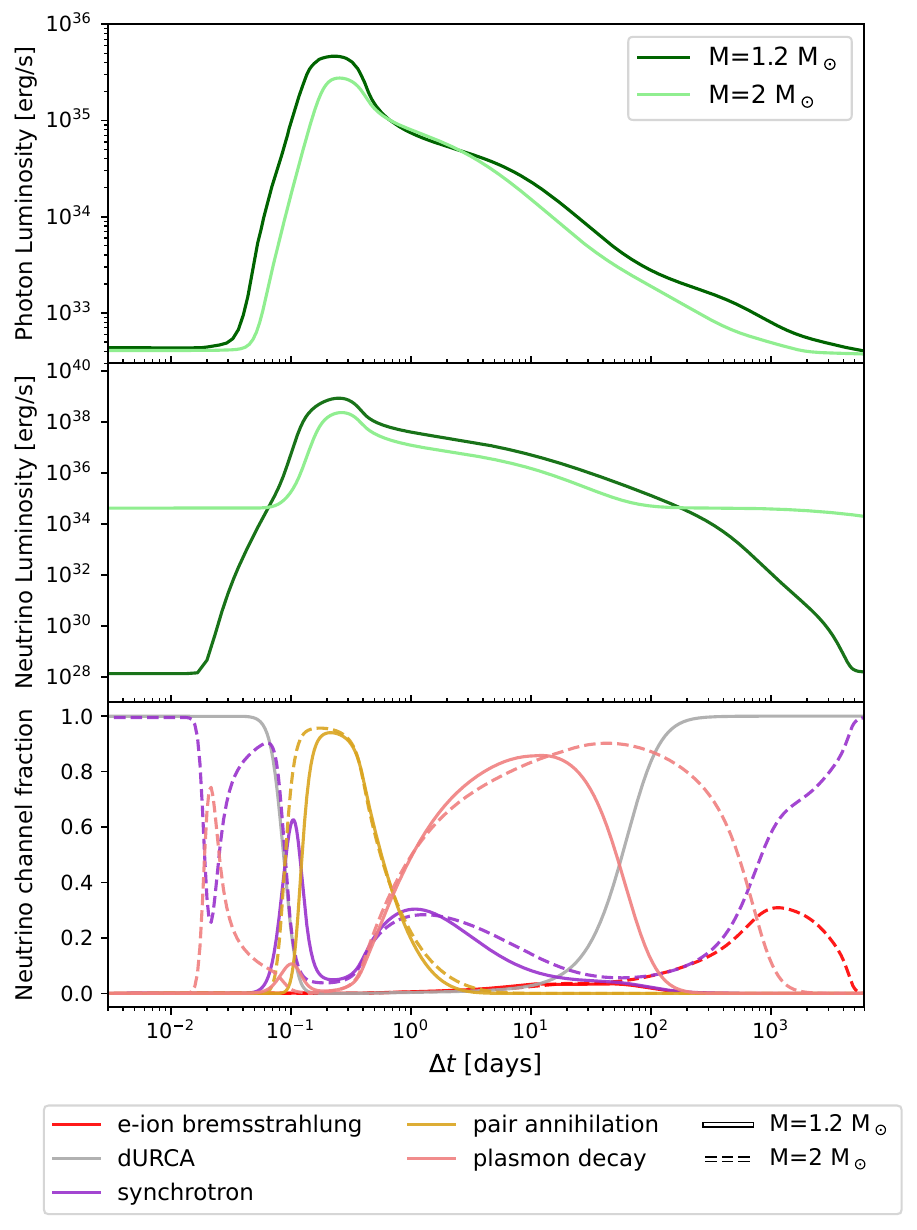}
    \caption{Photon and neutrino luminosity for two cases with heat injection down to $z_2=600\,$m for two NSs with different masses at the opposite ends of the possible values for NSs.
    The initial states were chosen in order to have the same quiescence photon luminosity (though the neutrino one is different). In the last panel, the fraction of the total neutrino luminosity due to the active neutrino reactions is displayed. The dominant cooling channel in the quiescence (background state) is different in the two cases, with the more massive star undergoing fast cooling via direct URCA reaction (gray line).
    }
    \label{fig:durca}
\end{figure}

Thus far, we considered as a background state a standard $1.4\,$M$_\odot$ NS. To study the effect of this parameter, in Fig.\ \ref{fig:durca} we compare the evolution of an outburst akin to the one studied in the previous cases ($E_\text{inj}=10^{43}\,$erg, $t_\text{inj}=10\,$h, $R=3\,$km, $z_\text{out}=5\,$m, $z_\text{in}=600\,$m) but for NSs having masses of 1.2 and 2 solar masses, at the upper and lower end of the values allowed by the EoS. In particular, that the latter case allows for fast cooling reactions, which for the BSk24 EoS are allowed above the critical mass $M_\text{dUrca}=1.595\,$\msun.
In order to better compare the ensuing cooling curves, we chose two states having the same photon luminosity in quiescence, thus starting the injection at $t_0=42$ and 120\,kyr respectively--though, since the evolution of the magnetic field and the corresponding Ohmic heating is not considered, these values do not have a strict physical meaning. The overall shape of the curves is remarkably similar, with several differences in the local slope of the cooling that can be associated to the different neutrino emission channels, becoming dominant at slightly different times, as displayed in the bottom panel. Hence, contrary to what happens for the long term cooling, the mass is not a crucial parameter in the cooling of the hotspot. This is due to the fact that the bulk of the mass, and thus the mass difference, is found in the core, whereas the outburst unfolds within the crust, a region that only has marginal differences in the two cases. The same principle can be applied to different equations of state: whereas the core composition is still very much unknown and varies widely among models, the crustal one is comparatively better understood, and the cooling of a crustal hotspot is not a good proxy for distinguishing between them.

\section{Conclusions}\label{sec:conclusions}

In this work, we presented an updated framework for the description of magnetar outbursts as short-term cooling events. 
In particular, we studied how the evolution of the luminosity of an outburst can put constraints on the physical parameters characterising the heat injection mechanism, whose precise origin remains elusive to this day.

Our simulations employ a state of the art framework to model the microphysical processes in the crust, alongside a self-consistent treatment of the outer boundary condition for the temperature evolution. Namely, we extended the integration domain of our simulation to include regions of low density, $\rho\approx10^7\,$\gcc, which required the use of magnetised envelope models that are thinner than what normally used in cooling simulations as a boundary condition. They can thus be used to study shorter timescales without breaking the plane-parallel approximation. With this setup, we could obtain the peak luminosity values observed in magnetars in outburst, $\lesssim10^{36}\,$erg/s, from a heated patch having a realistic emission radius ($\lesssim3\,$km). This result holds even when using envelope models with different chemical compositions (i.e.\ not just in a light-element scenario), which actually turns out not to be a crucial parameter, as neutrino emission is the limiting factor of the maximum luminosity achievable.

We studied the effect of several parameters related to the initial shape of the heated region, finding an overall similar behaviour, especially in terms of the total timescale of the event. On the other hand, many factors determine the specific features in the shape of the cooling curve. The initial depth of the heat deposition tracks the heat diffusion time up to the surface, and the slope of the photon luminosity curve acts as an indicator of the weak processes dominating the neutrino emission at a given time.

In order to account for the observed luminosity, we find that at least some part of the heating should happen in the very shallow layers of the crust ($\rho\lesssim10^8\,$\gcc), leaving open both the possibilities of an internal release at low density or a magnetospheric origin. On the other hand, constraints on the maximum depth that the heat can reach (or originate from) may come from the study of the late phases of the outbursts, when the hot spot has cooled down considerably. Still, most magnetars are already quite hot in their quiescence state ($\gtrsim 10^{33}\,$erg/s), so that the cooling of the hotspot in the late phases may be difficult to detect against the background emission. This makes highlights the importance to study the so-called low-field magnetars, a small class of objects that shows magnetar-like outbursts while at the same time exhibiting luminosity and spin-down field values akin to those of ordinary pulsars \citep{rea10, 2012ApJ...754...27R}. A continued monitoring of the cooling of their outbursts well after the peak, even for several years, may prove key to constraining the depth at which heat is liberated when the outburst is triggered, and hence the heating mechanism itself. However, the small number of these sources and the difficulties inherent in gathering long-exposure data data of dim sources that require long exposure times makes such detailed studies rather challenging.

Another factor rendering low-field magnetars particularly interesting is that they are expected to actually host ultra-strong fields, as strong as $\approx10^{16}\,$G, in small scale structure and/or a strong toroidal component buried under their surface \citep[e.g.][]{2011ApJ...740..105T, 2013Natur.500..312T, 2025NatAs.tmp...48I}. This could give rise to extremely large heat depositions, which, as in the case shown in Fig.\ \ref{fig:rin}, could display characteristic late-time features. 

The timescale over which the heating mechanism operates is another quantity that manifests itself directly in the shape of the luminosity curve, as it is well traced by the delay between the trigger and the onset of the luminosity descent. However, this implies that in order to characterise this parameter it would be necessary to have data about the outburst rise phase, or at the very least a stringent limit on its duration. Unfortunately, no information of this kind is currently available. New missions like \textit{Einstein Probe} \citep{Yuan2025} monitoring the X-ray sky with high cadence and sensitivity are going to provide fundamental information in the upcoming years, with the potential of putting the most stringent constraints to date to the trigger mechanism.

A key aspect of magnetar outbursts that we could not explore in the present study is how different magnetic field strengths and configurations affect the evolution of the event. We did test other field configurations, including some with several multipolar components structures as well as core-threading structures \citep[as in][]{2013MNRAS.433.2445A}, but found no significant difference with respect to the runs presented in this work. This is because the luminosity curves in the current 2D setup are affected by the field in the proximity of the pole and in the outer layers, where the deviation from a dipole cannot be but marginal in axial symmetry. Much in the same way, we tested different magnetar-like fields strengths without finding substantial differences with respect to the runs at $10^{14}\,$G presented above.
Nevertheless, the field structure has been shown to be a crucial factor using 3D codes: in particular, \citet{2022ApJ...936...99D} showed that if the heat is injected in a region where the field has a large component parallel to the surface (in contrast to the polar region considered in this work) the duration of the cooling phase can increase significantly due to the additional magnetic insulation. This may explain the prolonged enhanced brightness of some sources, such as {SGR\,J1745$-$2900} \citep[the so-called Galactic Centre magnetar,][]{2020ApJ...894..159R}. Moreover, the evolution of the field may play a role even on smaller timescales, e.g.\ in case of the presence of local instabilities (possibly related to the outburst trigger) or thermo-magnetic effects related to the large temperature gradient between the hotspot and the rest of the crust \citep[e.g.][]{2024A&A...690A.117G}. We defer the study of these effects to future work, employing a 3D code informed by our present 2D results.

Furthermore, in this work we restricted ourselves to the study of integrated quantities, rather than addressing the evolution of the observed spectral parameters, such as temperature and radius of the hotspot. This is due to the fact that these values are affected not only by the cooling of the hotspot itself, but also by other factors like the viewing geometry of the star, general relativistic effects, interstellar absorption and instrumental biases \citep[e.g.][]{2014MNRAS.443...31V}. The study of these parameters and how this information can open new windows in constraining the geometry of the surface of a magnetar will be addressed in future work.

\begin{acknowledgements}
      We thank A.\ Yu.\ Potekhin for discussion.
      DDG is supported by a Juan de la Cierva fellowship (JDC2023-052264-I). DDG and NR are supported by the European Research Council (ERC) via the Consolidator Grant “MAGNESIA” (No. 817661) and the Proof of Concept ``DeepSpacePulse" (No. 101189496). DDG, NR and KK are supported by the Catalan grant SGR2021-01269 (PI: Graber/Rea), and by the program Unidad de Excelencia Maria de Maeztu CEX2020-001058-M.6953. FCZ is supported by a Ramón y Cajal fellowship (grant650 agreement RYC2021-030888-I). Part of the simulations presented in this paper have been operated in collaboration with the Port d’Informació Científica (PIC) data center. PIC is maintained through a collaboration agreement between the Institut de Física d’Altes Energies (IFAE) and the Centro de Investigaciones Energéticas, Medioambientales y Tecnológicas (CIEMAT). 
\end{acknowledgements}

\bibliographystyle{aa} 
\bibliography{biblio}

\end{document}